\documentclass[journal]{IEEEtran}\usepackage{booktabs}
\usepackage{supertabular}
\usepackage[table,xcdraw]{xcolor}
\usepackage{adjustbox}
\usepackage{longtable} 
\usepackage{makecell, longtable}

\usepackage{stfloats}

\usepackage[skip=1ex]{caption}
\usepackage[notransparent]{svg}
\usepackage{booktabs}
\usepackage{multirow}
\usepackage{subcaption}
\usepackage[normalem]{ulem}
\usepackage{csquotes}
\usepackage{amsmath,amssymb,mathtools}
\usepackage{bm}
\usepackage{cite}
\usepackage{url}
\usepackage[separate-uncertainty = true]{siunitx}
\begin{document}

\title{A Systematic Review on Custom Data Gloves}

\author{Valerio~Belcamino, Alessandro~Carfì, Fulvio~Mastrogiovanni
\thanks{V. Belcamino, A. Carfì and F. Mastrogiovanni are with TheEngineRoom, Department of Informatics, Bioengineering, Robotics, and Systems Engineering, University of Genoa, Via Opera Pia 13, 16145, Genoa, Italy.}
\thanks{© 2024 IEEE.  Personal use of this material is permitted.  Permission from IEEE must be obtained for all other uses, in any current or future media, including reprinting/republishing this material for advertising or promotional purposes, creating new collective works, for resale or redistribution to servers or lists, or reuse of any copyrighted component of this work in other works.}}

\markboth{IEEE Transactions On Human Machine Systems}%
{Shell \MakeLowercase{\textit{et al.}}: Bare Demo of IEEEtran.cls for IEEE Journals}

\maketitle

\begin{abstract}

Hands are a fundamental tool humans use to interact with the environment and objects. 
Through hand motions, we can obtain information about the shape and materials of the surfaces we touch, modify our surroundings by interacting with objects, manipulate objects and tools, or communicate with other people by leveraging the power of gestures.
For these reasons, sensorized gloves, which can collect information about hand motions and interactions, have been of interest since the 1980s in various fields, such as Human-Machine Interaction (HMI) and the analysis and control of human motions.

Over the last 40 years, research in this field explored different technological approaches and contributed to the popularity of wearable custom and commercial products targeting hand sensorization. 
Despite a positive research trend, these instruments are not widespread yet outside research environments and devices aimed at research are often \textit{ad hoc} solutions with a low chance of being reused. 
This paper aims to provide a systematic literature review for custom gloves to analyze their main characteristics and critical issues, from the type and number of sensors to the limitations due to device encumbrance. 
The collection of this information lays the foundation for a standardization process necessary for future breakthroughs in this research field.
\end{abstract}

\begin{IEEEkeywords} 
Data Gloves, In-Hand Manipulation, Human Machine Interaction, HMI, Systematic Review.
\end{IEEEkeywords}

\IEEEpeerreviewmaketitle


\section{Introduction}  
\label{sec:Introduction}
Hands are complex and versatile limbs integrating advanced sensing and motion capabilities crucial for many daily activities, e.g., grasping, manipulation, and communication via gestures. 
Human hands are peculiar body parts where two senses, namely proprioception and touch, are closely affected by each other.

\begin{figure}
\begin{subfigure}{0.238\textwidth}
    \includegraphics[width=\linewidth]{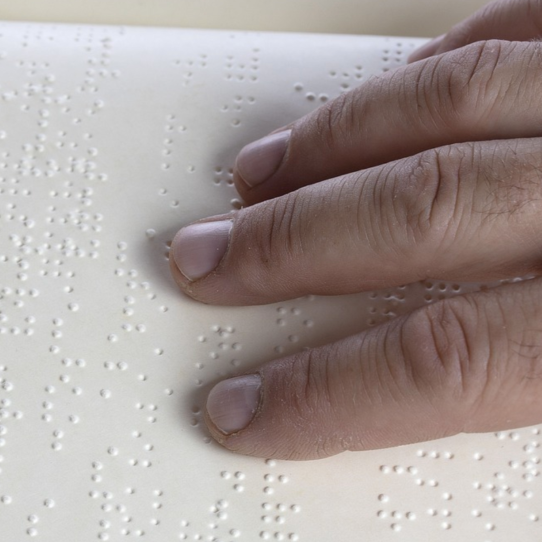}
    \par\medskip
    \includegraphics[width=\linewidth]{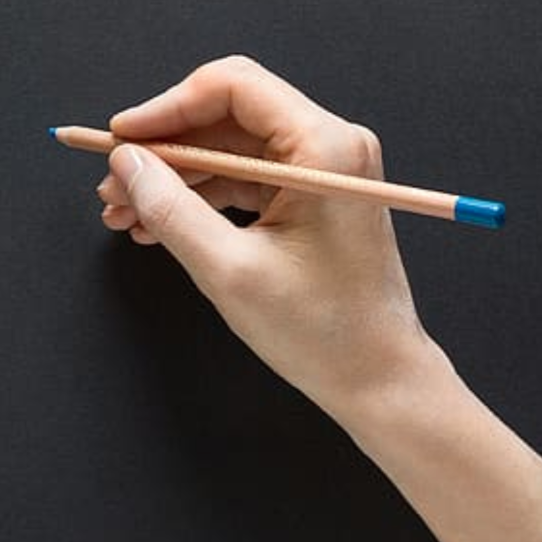}
\end{subfigure}
\hspace*{\fill}
\begin{subfigure}{0.238\textwidth}
    \includegraphics[width=\linewidth]{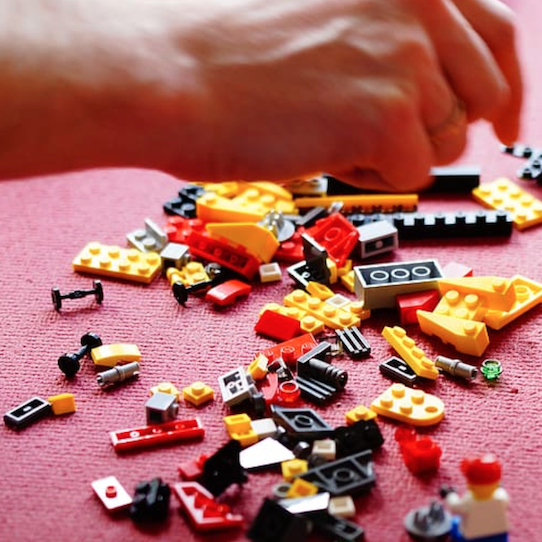}
    \par\medskip
    \includegraphics[width=\linewidth]{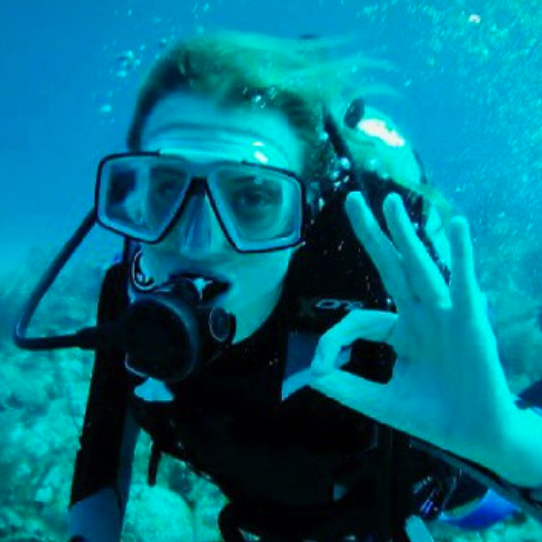}
\end{subfigure}  
\caption{Hands are of the utmost importance for a variety of human behaviors, such as sensing, manipulation, tool use, or communication.}
\label{fig:four-images-subcaption}
\end{figure}

In general, proprioception relates to estimating one's motion and posture.

Hands are tightly coupled with the brain and the nervous system, making them of the utmost importance for investigating traits of human behaviour, such as those related to motor control and the associated cognitive processes.

From an application-oriented standpoint, hands are likely the preferred physical medium enabling human-machine interaction, e.g., to use interfaces such as touchscreens or virtual reality controllers\cite{sim2021low, tsai2021unity}.
Therefore, a reasoned analysis of hand structure and purposeful usage is relevant in many research fields, for example, to replicate a synthetic surrogate of human perception traits.
To this aim, researchers need specifically designed tools to acquire meaningful data about hand motions and poses.

Studies in hand motion analysis can be categorized into two classes based on the adopted sensing modality, i.e., image-based and non-image-based \cite{carfi2021gesture}.
Approaches belonging to the first class rely on suitably located cameras to collect information about hand motions and, in some cases, may require a human to wear \textit{markers} to facilitate tracking, e.g., coloured gloves \cite{wang2009real} or passive infrared spheres \cite{han2018online}. 
Instead, approaches from the second class usually leverage sensors worn by humans, e.g., inertial measurement units (IMU) \cite{dong2021dynamic} or flex sensors \cite{aw2022data}. 
For these reasons, we will refer to the two classes, respectively, with the more technology-oriented terms vision- and wearable-based.
Whilst vision-based solutions are relatively simple to deploy since they usually rely on a single camera, but data are related to a specific, fixed point of view, wearable-based solutions usually require the use of multiple sensors (since their \textit{sensing space} is localized) thereby resulting in a more complex integration. 
For this reason, vision-based sensing is susceptible to occlusions and varying light conditions, which are inevitable drawbacks when observing human motions, whereas wearable sensors can be bulky and impair the hand's natural motion but allow one to perform motions virtually anywhere. 

Conversely, to study how hands physically interact with the environment and the objects therein in contact regimes \cite{Seminaraetal2023}, most technological solutions require the use of tactile sensors in direct contact with the hand \cite{wang2021grasping}\cite{biju2022sensor}, i.e., the sensors should be worn on the hand or otherwise embedded in the objects a human interacts with. 

Considering these two aspects as a whole, the use of wearable devices enables a complete perception of hand motions, poses, and the interactions between the hands and the environment\cite{liao2022plasma}\cite{steffen2020wearable}\cite{kerner2022wearable}\cite{zheng2022human}. 
As a further additional value, wearable devices nowadays find application in human-machine interaction since they can provide haptic feedback to their users, for example, using vibrating motors\cite{motors}, air chambers\cite{chamber}, or variable resistance motors\cite{exosc}.
Therefore, wearables are now a widely adopted solution to collecting data originating from human hand motions. 
Data gloves represent a popular solution for hand sensorization. 
The term \textit{data glove} describes a generic hand-worn device integrating one or more sensors for hand sensorization. 
Research activities on data gloves have a long history. 
In 1982, Zimmerman filed a patent application for a data glove using flexible optical sensors to measure finger bending\cite{zimm}. 
Later, together with Lanier, Zimmerman developed the Power Glove\cite{lani}, which includes ultrasonic and magnetic sensors to track hand motions and poses. 
Since then, the adoption of data gloves has grown over time, especially for research applications. 
To date, a wide range of data gloves is available in the full spectrum, from commercial solutions to custom designs.
Plug-and-play devices are available for sale, already calibrated, and ready out-of-the-box for immediate use.
Many commercial models with different sensor configurations are available: 
5DT Data Glove Ultra\footnote{\url{https://5dt.com/5dt-data-glove-ultra/}}, 
Hi5 VR Glove\footnote{\url{https://hi5vrglove.com/}}, and 
Manus Prime II\footnote{\url{https://manus-meta.com/}}, to name a few. 
Although these solutions are easy to use, they are rather expensive and lack the required flexibility to support extensive studies. 
On the one hand, as highlighted by Rodriguez \textit{et al}. \cite{caeiro}, commercial data gloves are characterized by an average price of over one thousand dollars and tactile sensing is rarely considered, i.e., only 7 out of the 30 data gloves considered in their study is a form of tactile sensing present. 
On the other hand, custom solutions provide maximum flexibility while maintaining costs low. However, their development is time-consuming and requires diverse expertise, e.g., in electronics design, firmware and software development, and ergonomics. 
Nevertheless, various research groups use custom data glove solutions. 
Despite a strong research interest, new works in the literature rarely reuse old designs but prefer to design their data gloves from scratch. 
We argue that this problem emerges from a lack of standards in the whole data glove technology stack. 

In order to address all of these challenges, it is timely and necessary to study existing data glove designs from the perspective of their hardware characteristics, the information they collect, and their target applications.
Apart from a few exceptions, and despite the unquestionable relevance of the issues discussed above, the specialized literature lacks a comprehensive review with a clear focus on custom data gloves and their design.
Dipietro \textit{et al}.\cite{dipi} presents essential information on the purpose, function, and main characteristics of existing, for the most part, commercial data gloves. 
Rodriguez \textit{et al}.\cite{caeiro}, following the PRISMA guidelines\cite{prisma}, presents a systematic review of the use and design of data gloves, but such review is limited to commercial products. 
Furthermore, different review studies are limited to specific application fields, e.g., virtual reality \cite{Perret}, diagnosis and prevention of rheumatoid arthritis\cite{artrite}, or robots teleoperation\cite{teleop}. 
Finally, other studies focus only on the interplay between wearable sensing and vision-based techniques \cite{multitrack,cvtrack}.

In this work, we address the limitations of existing literature reviews, and we present an in-depth systematic literature review for custom data gloves. 
We analyze 177 journal articles describing trends in design and application choices. 
Furthermore, we provide insight into the most promising directions for future research. 
The remaining part of the paper is structured as follows. 
Section \ref{sec:Methodology} describes the systematic review process. 
In Section \ref{sec:Relevance of the Topic}, we discuss the significance of the problem for different application scenarios.
In Section \ref{sec:Problem Statement}, we present the hand-sensing problem and describe the most common solutions. 
Finally, Section \ref{sec:Results} is focused on the results of the systematic literature analysis according to the definitions provided in the previous Sections. 
Conclusions follow.


\section{Methodology}  
\label{sec:Methodology}
\begin{figure}[t]
\includegraphics[width=1\columnwidth]{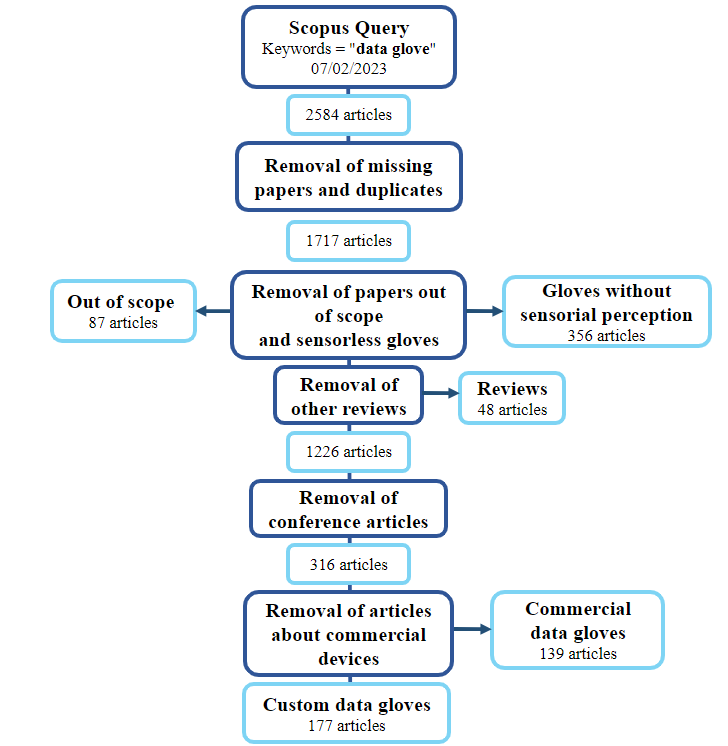}
\caption{
The selection criteria of the systematic review process. 
Each stage is associated with the used filtering criterion and the corresponding number of remaining articles.}
\label{fig:workflow}
\end{figure}

For our systematic literature review, we followed the well-known PRISMA guidelines \cite{prisma}. 
The articles considered in our analysis result from the filtering process shown in Fig. \ref{fig:workflow} on a query to the Scopus database performed on February 7, 2023. 
The query is as follows:

\begin{footnotesize}
\begin{verbatim}
    TITLE-ABS-KEY(data glove)
    AND (
        LIMIT-TO ( LANGUAGE, “English" ) 
    )  
    AND(
        LIMIT-TO ( SUBJAREA, “COMP" ) 
        OR
        LIMIT-TO ( SUBJAREA, “ENGI" ) 
    ) 
\end{verbatim}
\end{footnotesize}

The selected query limits the research to papers written in English (label: \textit{English}) in the computer science (label: \textit{COMP}) or engineering (label: \textit{ENGI}) fields containing the keyword \textit{data glove}. 
We decided to limit our search to the computer science and engineering fields since we want to focus on works addressing software and hardware design choices. 
Our query returned 2584 articles, and we processed all these articles to remove duplicates and those with no accessible full text or out of the review scope, i.e., those articles not addressing the sensing of hand motions.

After the preliminary filtering, all remaining 1717 articles describe contributions regarding sensing hand motions and poses. 
Next, we removed all the articles regarding sensorless gloves. This stage of the selection filtered out 356 articles. These articles mainly describe works in which coloured gloves are used to perform hand tracking through colour segmentation tasks or where the gloves are the support for motion capture markers. Additionally, all the review articles have been discarded from the systematic selection as they do not describe original research on custom data gloves. Among the 48 identified reviews, only 8 were relevant, and we discussed them in the Introduction of our paper.

We narrowed the review by focusing on journal articles and removing all the other types of publications. This filtering allows us to avoid considering different iterations of the same study, and it is based on the assumption that final results are typically published in journals. We classified the remaining 316 journal articles according to the following categories:

\begin{itemize}
\item
\textbf{Contribution}:
This category describes the main contributions of the paper. 
Articles can describe applications of custom or commercial data gloves; we define these as \textit{application articles}. 
In these papers, little to no description of the hardware is present. 

On the contrary, \textit{hardware articles} are those in which the data glove design is the main contribution.
It is important to remark that hardware articles may also propose one or more applications of the device they describe.    
\item 
\textbf{Availability}: 
With this category, we discriminate between articles in which \textit{commercial} gloves are employed, i.e., those identified by a model name and a manufacturer, and those in which \textit{custom} ones are present, i.e., data gloved designed and developed from scratch by the authors.
\item
\textbf{Type}: 
In this category, we divide articles based on the functionalities of the data gloves that are introduced or discussed.
In particular, we distinguish three types of data gloves, namely \textit{active}, \textit{passive}, and \textit{hybrid}. 
Active devices include in their design actuators, e.g., vibration motors \cite{motors} or air chambers \cite{chamber}, to provide tactile or pressure feedback to their users; passive devices only integrate sensors for collecting information from the hand motions and usage; hybrid data gloves combine both sensing and feedback.
\item
\textbf{Sensor}: 
This category describes the sensor types embedded in the device, if any.
\item 
\textbf{Application}: 
This category represents the application field considered in the article. 
After a careful analysis of the literature, we identified three application fields we focus on, which we will describe in Section \ref{sec:Relevance of the Topic}.
\end{itemize}

\begin{figure}[t]
\includegraphics[width=1.00\columnwidth]{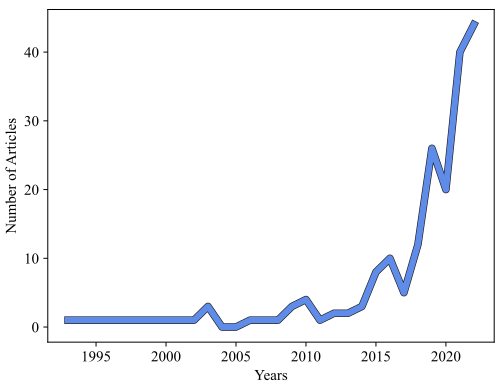}
\centering
\caption{The trend in the number of journal articles about sensorized custom data gloves over time.}
\label{fig:narticles}
\end{figure}

We used the classification results for an additional filtering phase. 
Since the scope of this article is to study data gloves for the sensing of hand motions, we discarded all articles describing \textit{active} data gloves. 

Finally, of the remaining 316 journal articles, we disregarded those using commercial products. 
The outcome of the filtering procedure is a group of 177 journal articles regarding passive or hybrid custom data gloves.


\begin{figure}[t]
  \includegraphics[width=\columnwidth]{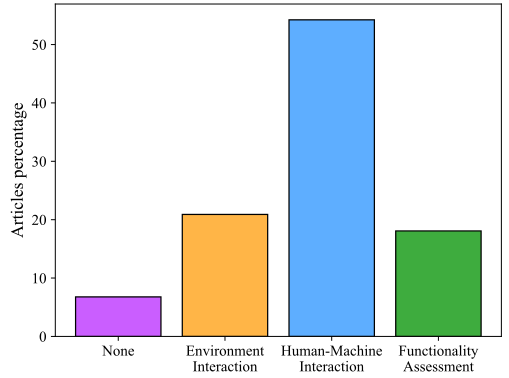}
  \centering
  \caption{The percentage of articles considering each application field on the total number of selected articles. A publication can only belong to a single category and None has been used for articles without a specific application.}
  \label{fig:applications}
\end{figure}

\section{Relevance of the Topic}  
\label{sec:Relevance of the Topic}
As mentioned in the Introduction, tracking hand movements, poses, and the forces exchanged between hands and the environment, including objects, during physical interaction has been the main topic of many studies, given their great relevance in different research fields. 
According to our review results, the literature about custom data gloves started in the early 80s and saw a significant increase in publications in the last ten years, as shown in Fig. \ref{fig:narticles}. 


A possible explanation for this rapid growth may be the increasing availability of miniaturized sensors at affordable prices.
Furthermore, the adoption of custom devices allows for adapting the hardware design to the application of interest, for example by choosing the number and type of sensors as well as the material and shape of the glove. 
Such flexibility is one of the main advantages of custom over commercial solutions and possibly one of the main driving factors of research activities in this field.
The increasing volume of the corpus of research on custom data gloves also results from the interest in hand-related studies that emerged in different research fields. 
As anticipated in the previous Section, during the analysis of the current literature we identified three major research fields which are particularly active in the adoption of custom data gloves, namely Functionality Assessment, Environment Interaction, and Human-Machine Interaction. 
These areas have been identified simply by clustering articles based on their stated research objectives.
Studies in Functionality Assessment mainly use custom data gloves to measure the motion of the hand for diagnostic purposes \cite{CROOK199870, Ku2003ADG, Tanakaeii, lorussi2016wearable, saleh2016monitoring, lin2017data, asadipour, 8118157, elwahsh2022new, stornelli201910} or to track the performance of a rehabilitation \cite{hoda2018predicting, guo2021design, davarzani2021design, elksass2022anti, sarwat2021design,  maddahi2022roboethics, 9531064}, and therefore the possibility of custom design choices to explore possible therapeutic parameters plays a major role.
Environment Interaction applications include broad studies focusing on the movements and forces generated by the hand during contact regimes\cite{ amirouch, Kingra, falco2012data, fujiwara2013evaluation, Hazmanmu, ayodele2021grasp, Ayodeleemm, maitre2021object, melnyk2019physical, shahrampour2017online, kao2018novel}. 
Goals include, for instance, the analysis of ergonomics-related measures, or the use of certain materials in product design.
Finally, for Human-Machine Interaction, we can distinguish between studies targeting implicit or explicit communication. 
while the first class includes approaches aimed at recognizing human actions\cite{malawski2018system}, for example, to assess the activities of daily living, studies belonging to the second class typically adopt data gloves as input devices to teleoperate robots\cite{Dalleysky, girovsky2015robotic, Yahyama, ullah2019single, lishiui, syed2019flex, setiawan2020grasp,  huang2022virtual, fangbin, yudhana2022performance, muezzinoglu2021intelligent} or control software applications\cite{kim2019performance, jiang2020fiber, noh2020development, mraz2003platform, Gallo2010, linbre, lin2022interactive, wang2022design, 9969542, fu2021design, 15948}, e.g., via gestures\cite{ zhang2019cooperative, huang2019tracing, mehra2019gesture, chuang2019continuous, 8701582, chitra, pan2020wireless, jhamb2020wireless, shin2021hand,  yuan2020hand, yuan2021gesture, 9716737, li2021real, ahmed2021real, piskozub2022reducing, calado2022geometric, hekmat2022map, 9760170} or communication with sign languages\cite{4154662, shukor2015new, kim2016recognition, galka2016inertial, li2016sign, saldana2018recognition, Puasaga, shweta2019user, alrubayi2021pattern, 9830849, dias2023comparison, calado2021toward, zhang2022sign, simoes2022instrumented, Ahmedmo}. 

It is worth noting that this simple classification does not have the pretence of being exhaustive, and its main objective is to describe custom data gloves' relevancy in different application fields, see Fig. \ref{fig:applications}. 
We should also specify that not all the reviewed articles target a specific application.
Instead, many of them present only a hardware architecture \cite{ 1604113, 69485, yuen2014proprioceptive, zuruzi2017towards, zhang2019textile, 7894, song2020mechanically,  zhang2020highly, liyatao}. 
These articles are identified by a \textit{None} label in Fig. \ref{fig:applications}. 
The analysis of article distribution over the different application fields will be used later in our review to describe data glove design trends as observed in the literature. 

Despite the significant research interest in custom data gloves, the literature is fragmented and lacks standardization efforts, which results in low reusability of published results. 
Most data gloves in the literature can not be easily reproduced because of poor hardware and software description. 
For these reasons, researchers often prefer to design and manufacture a new data glove without reusing previous -- typically very similar -- solutions. 
Given these conditions, our literature review is the first step towards a standardization of the research field, a topic of increasing importance because of the vast interest in this research topic.


\begin{figure}[!t]
\includegraphics[width=\columnwidth]{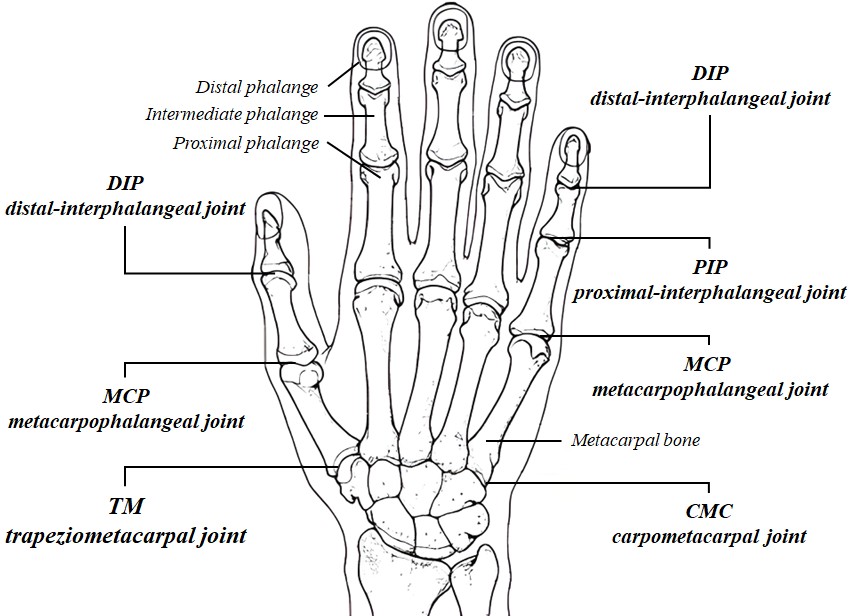}
\centering
\caption{A human hand kinematic model from \cite{phdthesis} considering all DoFs except for the wrist.}
\label{fig:27DOF}
\end{figure}
\section{Problem Statement}  
\label{sec:Problem Statement}


\subsection{Rationale and Scope}

Humans have highly developed perception capabilities resulting from the interplay of all senses, allowing them to collect via receptors and integrate into neural pathways information from the surrounding environment. 
In doing this, hands obviously play a crucial role since they are key means for both sensing and action unfolding \cite{Seminaraetal2023}.
Proprioception and touch are fundamental for humans to estimate \textit{intrinsic} information, for example, the pose of fingers or the hand as a whole\cite{wu2021development, nassour2020robust, zhang2020static, liu2019new, zhang2019high, chacon2018development, da2011fbg}, as well as \textit{extrinsic} information, such as contact locations and shapes\cite{cazacu2021position, liao2018hierarchically}, or the distribution of external mechanical loads \cite{proff}. 
Intrinsic and extrinsic information gathered by hands extend and integrate those from other essential senses, e.g., sight and hearing. The interplay with other senses, especially with sight, is fundamental in everyday life activities and for actions requiring a high level of eye-hand coordination \cite{handeye} like catching an object mid-air or the manipulation of extremely tiny objects \cite{tinyobj}. Furthermore, in some cases, hands' senses can even substitute other senses, as it happens for blind people using touch to read \cite{braille} or for the localization and identification of objects in low-light environments \cite{tactilexplo}. 

Accordingly, the primary goal of hand sensorization, and therefore of custom data gloves, is to extract intrinsic and extrinsic information from hand motions and their physical interaction with the environment. 
Solutions for hand perception can vary to a large extent according to the adopted sensing strategy.
However, current approaches follow the typical computational perception pipeline characterized by the two phases of sensing (related to the mechanical transduction of physical phenomena) and processing (aimed at extracting semantically grounded information from raw data).

Although a general treatment of hand sensorization is more complex, we limit the scope of our study primarily to the perception of hand motions and, to a lesser extent, to the detection of contacts.
These are those hand-sensing aspects that received more attention in the literature about custom data gloves and for which an accurate treatment would greatly benefit research and technological development. 
Therefore, the following Sections describe the challenges associated with hand-related motions and their perception of touch.

\subsection{Modelling and Detection of Hand Motions}

To replicate the proprioception capabilities of human hands, it is necessary to estimate both the hand pose (i.e., the whole hand position and orientation) and its kinematic state (i.e., the mutual configurations of fingers, palm, and wrist). 
This result can be achieved using instruments often referred to as tracking devices and exploiting different sensing technologies, such as vision or wearable sensors. 
Obviously enough, the collected sensory data vary according to the adopted sensing strategy. 
However, in the literature, there is reasonable consensus on a general approach whereby raw data are processed to reconstruct both the hand pose and its internal state. 
The latter is made up of all the hand joints and links organized in a formal, morphological representation, and its evolution over time describes hand configurations. 
This representation allows us to use joint angle trajectories to describe complex actions, such as grasping- and manipulation-related motions.

Formalizing the hand kinematic model starts from obvious facts.
As described by Tubiana \textit{et al.} \cite{handskele}, the human hand has a thumb and four fingers, namely the index, middle, ring, and little fingers. 
Fingers consist of three bones called distal, intermediate, and proximal phalanges, connected by joints called metacarpophalangeal (MCP), proximal-interphalangeal (PIP), and distal-interphalangeal (DIP), as shown in Fig. \ref{fig:27DOF}. 
Furthermore, if we also consider metacarpal bones, an additional joint should be added to the kinematic structure of each finger, that is the trapeziometacarpal (TM) joints. 
The thumb differs from the other fingers because it has only two bones, called distal and proximal phalanges, and three joints, i.e., the TM, the MCP, and the DIP.
These articulations allow for multiple DoFs distributed as one for DIPs and PIPs, which can only perform bending and extension, and two for the MCPs, which can also perform adduction and abduction movements. 
Rath \cite{mcpdof} also assumes that the MCPs can rotate around their axis, yielding a better grip while grasping objects. 
However, this DoF is not directly controlled by muscles. 
The TM joint, which is only present in the thumb and connects it to the carpal bones, enables the opposition to the other fingers, enabling two additional DoFs. 
Lastly, the CMC joints could introduce more DoFs, but given their low mobility and position, they are never addressed in the literature related to hand sensorization. 
Considering only those DoFs humans can voluntarily control, we have described 21 DoFs so far, i.e., four for each finger and five for the thumb. 
Hossain and colleagues \cite{hossa} suggested that the total number of DoFs of the hand is 27, identifying the remaining 6 in the wrist. 
Three of these DoFs are related to flexion and extension, adduction and abduction, and pronation and supination, while the last three are related to translation in space. 
Although hand kinematics is quite complex, research studies can choose the number of DoFs to be monitored according to the target application. 
In fact, it is common to consider simplified versions of the kinematic model, ignoring specific fingers or joints, depending on the applications.

If we limit our analysis to wearable solutions for hand motion tracking, we can identify three different categories of sensors: flex sensors, IMUs, and magnetic trackers. Although all these sensors can track hand motions, they do it by measuring different physical quantities, which leads to different advantages and disadvantages from a technological standpoint.

Classical flex sensors are made of a conductive filament that changes its resistance in a way proportional to the bending amount \cite{luo2022heterogeneous, zakridesigning, duan2022highly, huang2022high, HanMiao}.
Other solutions use optical fibres \cite{fiber, fujiwaraer}, capacitors \cite{capa}, or conductive inks \cite{ink}. 
When sewn or knitted \cite{lee2021knitted} into a data glove and aligned with the joint to track, flex sensors can measure the amount of bending of the surface around the joint.
However, sensor measurements are not equivalent to the joint angle variation, and a calibration procedure is typically needed.
To this aim, sensor values corresponding to known or suitable angles are first collected, and then intermediate values can be obtained through linear interpolation. 
Obviously enough, one of the main issues with flex sensors is that, after prolonged use, materials may degrade, requiring periodic re-calibration. 
Furthermore, flex sensors can track one DoF at a time, making it challenging to design complex sensing configurations..

IMUs are a class of devices integrating accelerometers, gyroscopes, and often magnetometers. 
The physical quantities measured by an IMU are related to the linear acceleration, the angular velocity, and the magnetic field when the magnetometer is included. 
Furthermore, data provided by IMUs can be processed with data fusion algorithms\cite{choi2018development}, such as the complementary filter, for example, to estimate the sensor's orientation in space \cite{kalman}. 
Therefore, IMUs are typically used in pairs to estimate joint rotations by measuring the orientations of the pairwise connected links. 
In fact, assuming a shared reference frame obtained through calibration \cite{narvaez2018quaternion}, it is possible to compute the relative rotation between the two sensors and track up to 3 DoFs of the joint. 
The main disadvantage of this solution is that IMUs' orientation estimate suffers from drifting caused by the integration of the error over time and by fluctuations in the magnetic field, as shown by Guidolin \textit{et al.} \cite{9385684}.

Finally, magnetic trackers\cite{1413566, 9376979} use electromagnetic induction to estimate the position and orientation of a receiver with respect to an emitter such as an electromagnetic dipole field generator. 
Therefore, as with IMUs, it is possible to estimate a joint angle by positioning the receivers on the two consecutive links that it connects.
This technology does not have any drifting and occlusion issues. 
However, it is affected by the interference generated by certain materials when submerged in a magnetic field. 
Aluminium and copper are among the materials that should be avoided because the magnetic field generated by the emitter induces a current on them, that is, \textit{eddy currents} \cite{Ferreira1989}, generating an additional magnetic field that interferes with the main one \cite{magne}. 
This technology is widespread in commercial applications, and it is used by many companies to produce hand-tracking devices. 
The Flock of Birds by Ascension Technology Corporation and the Polhemus Fastrak\footnote{\url{https://polhemus.com/motion-tracking/all-trackers/fastrak}} are among the most well-known devices on the market exploiting this technology.

It is noteworthy that the sensor type is not the only aspect of interest when it comes to tracking hand motions since their number and positioning are equally significant features to consider during the design phase of custom data gloves. 
Furthermore, hand motion tracking is susceptible to an additional challenge, i.e., the alignment between sensors and tracked joints.
Even a small misalignment between sensors and fingers can result in possibly relevant errors in joints' angle estimation. 
In the literature, this is often referred to as sensor-to-segment calibration \cite{sensor_segment, sensor_segment2}, and can be addressed by following four different strategies:

\begin{enumerate}
\item

assumed alignment: the misalignment of the sensor with respect to the body segment is reduced by manual calibration. During this process, alignment between the two reference frames is enforced by proper positioning and orientation of the sensor on the segment \cite{s20113322};
\item
functional alignment: the misalignment between finger and sensor is estimated by measuring the orientation of the two reference frames during the execution of a series of  known calibration movements \cite{functional1,functional2};
\item
model-based: the rotation between the finger and the sensor is estimated by leveraging the constraint imposed by the hand kinematic model \cite{seel2012joint};
\item
augmented data: misalignment is corrected using an external device that measures the relative orientation between the sensor and the body segment. The most common systems employed in this type of calibration are RGB-D cameras and motion captures \cite{picerno2019upper,WEYGERS2021110781, jiang2019model};
\end{enumerate}

An additional factor to consider while designing wearable hand-tracking solutions is how sensors are mechanically fixed to the hand. 
Often, in the literature, sensors are assumed to be \textit{embedded} in a glove\cite{karlsson, MDLSP02, Tarchanidiskost}. 
However, solutions based on actual gloves are characterized by many drawbacks. 
Among these, the material can stretch, causing the uncoupling between sensors and hand motions, and it is difficult for gloves to adapt to users with different hand sizes. 
Therefore, a wrong choice of materials and size can negatively impact the ergonomics and sensing accuracy of the device. 
If a glove is too loose, it can yield unwanted sensor displacements, and conversely, if it is too tight, it can limit the fingers' range of motion. 
Alternatives to the glove as a sensing substrate, such as hand exoskeletons\cite{456879, You2001ALI}, rings\cite{zhou2022learning, bouzit}, and thimbles, exist in the literature. 
These approaches are expected to improve sensors' stability at the cost of increased bulkiness and setup time.

In some use cases, the requirement of a hand kinematic model to ground the hand pose estimate can be relaxed.
A few approaches posit that it is possible to recognize hand motions directly using raw data (i.e. accelerations, angular velocities and magnetic field values for IMUs and voltage for FSRs). In these cases, the sensor readings are generally used to train deep learning models, which can perform classification \cite{classification}, regression \cite{regression} or generation of synthetic movements \cite{lastrico2024effects}. The main models implemented in the literature are Long Short-Term Memory (LSTM) Networks \cite{lstm}, 1-dimensional Convolutional Neural Networks (1D-CNN) \cite{cnn}, Transformers \cite{transformer}, Generative Adversarial Networks (GANs) \cite{gan} and Variational Autoencoders (VAE) \cite{vae}. As for the application fields, gesture recognition is the most likely to adopt deep learning strategies as it is generally modelled as a classification task, which can be easily solved with data-driven approaches regardless of the kinematics model.
Nevertheless, it is possible in any scenario to create model-based classifiers that receive input from the estimated kinematic model angles. This strategy is commonly utilized when the goal is not only to recognize gestures but also to simulate hand motions.

\subsection{Modelling and Detecting Contact Phenomena}

Humans can perceive contact with objects in their surrounding environment leveraging the sense of touch.
Sensory receptors of different mechanical transduction capabilities can respond to exerted forces and vibrations originating from contacts and convey encoded information via neural pathways to the somatosensory cortex.
From a morphological perspective, the mechanoreceptors are characterized by a physical shape that allows them to sense different physical properties: 
Meissner corpuscles are sensitive to light touch and to the vibration of approximately 50 Hz, thereby being instrumental for slip detection, whereas Pacinian corpuscles detect rapid vibrations (of about {200-300} Hz) over large areas of the skin; 
Ruffini endings detect tensions, being therefore sensitive to skin stretches, whereas Merkel discs are sensitive to durative pressures and are therefore of the utmost importance to classify contact shapes.
Furthermore, temperature and pain can be detected via appropriate receptors, which can be useful to support quick motion reflexes related to safety and preservation \cite{Kandel2021}.
As explained in early work by Vallbo \textit{et al}. \cite{vallbo1984microstimulation}, among others, these mechanical receptors are present in various densities in the hands and, specifically in the case of touch, in the fingertips.
Such an evolutionary advantage, coupled with fine hand motions, allows hands to distinguish surface details, shapes, and other mechanical features of objects during manipulations and to identify the location and stability of contact points \cite{JohanssonFlanagan2009}.
Indeed, complex sensorimotor loops are believed to partake in such haptic capabilities, which are later integrated with visual processing in the brain \cite{Seminaraetal2023}. 

Data gloves aimed at acquiring tactile information must be equipped at least with sensors that can react to pressure distributions exerted on the hand while in contact with the environment. 
A typical solution adopted to measure such pressure distributions assumes the adoption of conductive materials whose resistance varies when their physical structure is subject to stress or strain due to external forces\cite{hang2020highly, zhang2020static, zhang2021master, li2020ultra}. 
If we focus on a high technology readiness level, the most common type of sensor aimed at that is the Force Sensing Resistor (FSR)\cite{Biju_2021}, which is built using conductive polymers.
FSRs are not the only solution, though. 
Similar technologies have been proposed in the past few years, such as piezoelectric crystals, optical fibres, strain gauges, conductive fluids, or capacitive sensors. 
These sensors have been used as transduction devices to study force distributions in grasping actions \cite{4694636}, they have been embedded into objects to estimate contact points during manipulation tasks \cite{9743777}, or have been organized into arrays mimicking a morphology inspired by human hands \cite{Wettelsetal2008}. 
Given the similar working principle, their calibration and the interpretation of measurements are subject to the same issues and procedures needed to operate flex sensors. 

Other technologies to detect contact forces have been explored in the Robotics-related literature, both for manipulation purposes and other robot body parts, but not mature enough for professional use or commercial exploitation.
These include soft pressure sensors able to distinguish force components along the three axes \cite{jammo}, which may be instrumental to detecting both normal and shear forces \cite{ Waskoetal2019}, as well as large-scale tactile sensors in the form of \textit{robot skin}, which can be useful to locate contacts or determine their shapes \cite{Dahiyaetal2010, Schmitzetal2011, Billardetal2013, Chengetal2019}. 
Although these types of sensors can provide a wide range of tactile-related information, they have been rarely employed in data gloves \cite{tact1}.

When it comes to custom data gloves for hand sensorization, or more in general to wearable devices, the present lack of popularity of advanced tactile sensors may be justified by many drawbacks hindering their adoption, namely their relatively high price, their inherent complexity, which encompasses the design of -- often flexible -- electronics \cite{Maiolinoetal2013b}, cabling \cite{Baglinietal2010} and real-time data processing \cite{Youssefietal2015}, material and manufacturing issues, which is related to their durability and long-term performance \cite{Loietal2013}, as well as their overall bulkiness, which makes it difficult to integrate them on small devices or mechanical structures, and significantly impairs the user's perception of contacts.

\begin{figure}[!t]
\centering
\includegraphics[width=0.8\columnwidth]{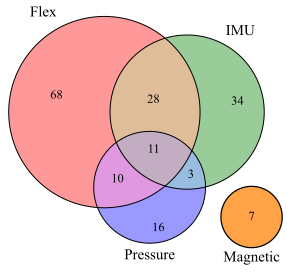}
\caption{
The number of papers mentioning each sensor category.
Flex sensors are employed by 68 articles, IMUs by 34, pressure sensors by 16, and magnetic sensors by 7 papers. 
Moreover, the Flex-IMU intersection counts 28 articles, IMU-pressure 3 and Flex-pressure 10. 
The remaining 11 articles include all the sensors.}
\label{fig:SensorType}
\end{figure}

\section{Results}  
\label{sec:Results}

In the previous Sections, we introduced the main ideas underlying custom data gloves, described the challenges they target, and analyzed their design issues. 
In light of this analysis, in this Section, we present the results of our literature review.

One of the most relevant aspects of custom data gloves is the hardware they integrate, which allows us to classify devices for hand sensorization according to their functionalities. 
As described in Section \ref{sec:Methodology}, we can distinguish between three types of data gloves, namely active, passive, and hybrid. 
However, we should point out that in our review process, we discarded all those works addressing active data gloves, that is, data gloves without sensing capabilities. 
Among the remaining articles, approximately only 10\% of the devices are hybrid, while the remaining 90\% are passive.
Hybrid gloves are adopted to provide feedback through exoskeletons, vibrating motors, and air chambers.
Trying to fit many actuators and sensors within the limited surface of the hand can often result in bulky solutions, which can impair user movements, eventually leading to poor ergonomic indicators. 
For these reasons, the usage of hybrid data gloves is restricted to specific scenarios, for example, in physiotherapy \cite{phi}.

\begin{figure}[!t]
\includegraphics[width=\columnwidth]{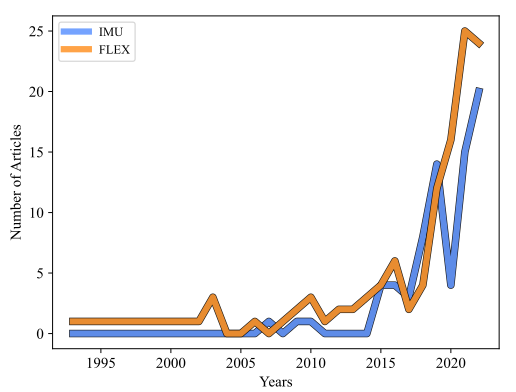}  \centering
\caption{A comparison between the number of articles mentioning IMUs and flex sensors over time.}
\label{fig:narticomp}
\end{figure}

Another fundamental distinction is related to the types of sensors integrated into the glove's structure. 
As discussed above, we identified four sensor types, i.e., IMUs, flex, and magnetic sensors to measure hand motion, and tactile sensors to characterize contact events. 
In Fig. \ref{fig:SensorType}, we provide a perspective on the literature that associates a set of dimensions to each sensor type, whose values are proportional to the number of articles adopting that particular technological solution. 
Intersections between sets represent those contributions to the literature using multiple sensor types.
It emerges that for articles considering a single sensor type, flex sensors are the most common (68 articles), followed by IMUs (34 articles) and pressure sensors (16 articles). 
Furthermore, 15.6\% of the articles employed a combination of IMUs and flex sensors, 1.6\% used IMUs and pressure sensors, 5.5\% used flex and pressure sensors, whereas 6.1\% are characterized by a setup making use of all sensor types. 
Magnetic trackers are only described in 7 publications and are never used in combination with other sensor types. 
For this reason, we did not consider them in our analysis. 
These results emphasize that most research interest in data gloves addressed hand motion tracking, while touch sensing has been limited to specific applications, for instance, in studies related to hand-object interaction. 

If we restrict our analysis to motion tracking, the predominance of flex sensors is evident. 
This technology is used in 66.6\% of the approaches, while IMUs are adopted in the remaining 33.3\%. 
This imbalance in the number of solutions adopting flex sensors may derive from their integration and simplicity of use. 
A single flex sensor can be positioned over a joint to generate an estimate of its angle with a simple calibration process. 
Instead, data originating from IMUs should be processed with a full computational pipeline employing data fusion algorithms for an accurate estimate of the sensor orientation \cite{estimation}. This approach requires pairs of sensory modules to track a single joint, i.e., sensors should be positioned on pairwise connected links with respect to the monitored joint in the kinematic chain. 
An advantage of IMUs over flex sensors is given by the number of DoF they can observe. 
Each couple of IMUs can estimate three DoF \cite{kall}, whereas flex sensors monitor one DoF each \cite{saggio2015resistive}.
Given this difference, IMUs are more suitable to monitor joints with multiple DoF, for instance, finger MCP, whereas flex sensors can be more successfully employed to measure joints characterized by a single DoF, for example, finger DIP and PIP. 
Fig. \ref{fig:narticomp} shows a comparison in the number of articles addressing IMUs and flex sensors over time.
We can observe how both categories follow the general growth trend reported in Fig. \ref{fig:narticles}.
In particular, flex sensors appear to be generally more popular, especially if we focus on older articles. Furthermore, both categories experienced sharp growth after 2017, leading them to over 20 journal articles a year after 2020.

\begin{figure*}[!t]
    \includegraphics[width=0.8\textwidth]{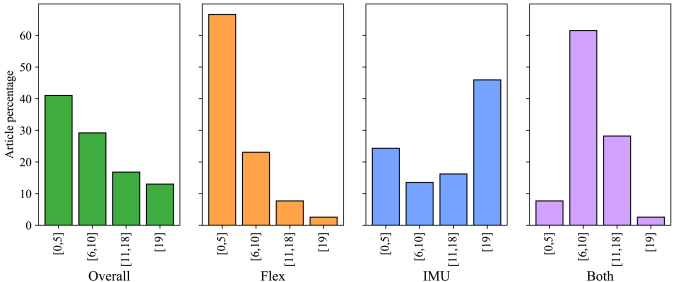}  \centering
  \caption{Comparison of the number of DoFs for each sensor type. The articles were classified according to the three classes introduced in this article.}
  \label{fig:newisto}
\end{figure*}

\begin{figure*}
\centering
\begin{subfigure}{.5\textwidth}
      \centering
      \includegraphics[width=\textwidth]{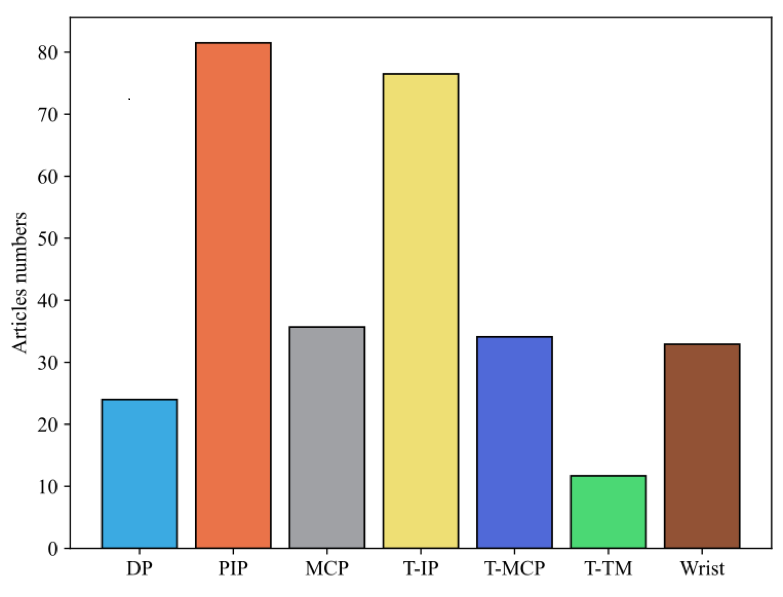}
\end{subfigure}%
\begin{subfigure}{.5\textwidth}
      \centering
      \includegraphics[width=0.65\textwidth]{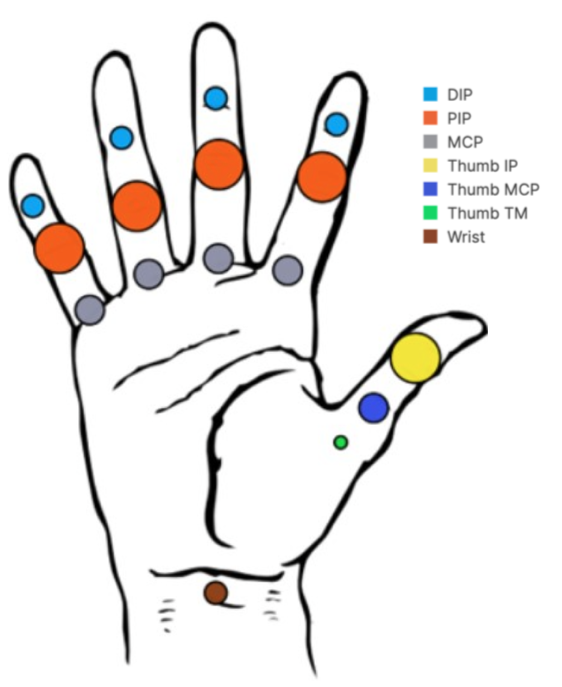}
\end{subfigure}
\caption{On the left, the percentage of articles mentioning each joint of the hand, including the wrist. On the right is the visual representation of this distribution.}
\label{fig:dofdist}
\end{figure*}

If we consider hand motion tracking, we defined four categories to discriminate data gloves based on the number of DoFs they can track. 
The four categories are \textit{basic finger tracking} (BSF, up to 5 DoFs), \textit{enhanced finger tracking} (EFH, 10 DoFs maximum), \textit{hand-wrist tracking} (HWT, up to 18 DoFs), and \textit{complex manipulation model} (CMM, more than 18 DoF). 
These categories and the corresponding DoF ranges were identified by studying the current state-of-the-art and clustering data gloves based on their applications. 
On the one hand, BSF data gloves can be used for the precise tracking of a single finger or with sensors distributed all over the hand for gesture recognition and for applications using simple human-machine interfaces\cite{fu2009user, fu2010building, chenpei, saini2015designing, kim2016improving, CheAni2016RealtimeFH, lu2018stretchable, javaid2018communication, umar2019occupational, lim2019development, revisore}. 
EFH data gloves constitute a fair trade-off since they can track most of the hand DoFs while the remaining ones can be estimated using kinematic constraints.
They can be used for simple hand motion studies. 
On the other hand, HWT data gloves usually do not limit the motion tracking to the hand but involve the wrist as well. 
This peculiarity allows for more complex studies about hand usage in hand-object interaction. 
Finally, CMM data gloves track almost all the DoFs of the hand\cite{lin2022design, cerqueira2022glove, bright2022low, 9741554, 9628050, ganjdanesh2017new, glauser2019interactive, linleee, 8762146}, are characterized by a more complex design and manufacturing, and their use is limited to advanced studies in the analysis of the motion of human hands. 

In Fig. \ref{fig:newisto}, we present the distribution of articles over all four categories and highlight the distribution difference according to the adopted sensors. 
It is noteworthy that most of the papers we review here do not explicitly specify the number of tracked DoFs.
We retro-engineer this number based on the type and the number of adopted sensors. 
Therefore, this analysis reflects the capability of the hardware explicitly mentioned in the reviewed papers but not necessarily how it had been used in the actual work. 
Therefore, to the best of our knowledge, we can observe that BSF data gloves are the most common solution (40\% of the articles), and popularity decreases as the number of tracked DoFs increases. 
From Fig. \ref{fig:newisto}, we can notice that flex sensors drive this trend, whereas IMU-based solutions seem to follow the opposite trend, having CMM as the most popular class.
Another interesting insight comes from the articles integrating IMUs and flex sensors, for which the EFT category is the more popular (61\%).

These results suggest that flex sensors are preferable for tracking low DoF numbers, whereas IMUs better suit applications needing a more precise hand kinematic model. 
This observation can be justified by considering the characteristics of the sensors analyzed in this article. 
Flex sensors are generally long and narrow strings that can track only one DoF at a time. 
With this technology, to measure multiple DoFs of a joint, it is necessary to position multiple sensors closely but along different axes. 
However, this operation should be accurate to prevent sensors from measuring the motion of other joints. 
Furthermore, the overlap of multiple flex sensors on the same joint could obstacle the free motion of the joint. 
Instead, only a couple of IMUs positioned on pairwise connected links across a joint are necessary to track its 3 DoFs. 
In addition, with three sensors placed on consecutive links of the kinematic chain, it is possible to track 6 DoFs, one for each ordered couple of IMUs. 
However, it is necessary to remark that assembling IMU-based data gloves is challenging since IMUs are typically bulkier and require more cables.

In addition to the number of sensors, we should consider their location on the hand. 
Depending on the application, some sensors are more important than others, and a careful choice of their location can have a massive impact on the overall system's effectiveness. 
Certain joints may also be difficult to track due to their shape, such as the TM of the thumb, or may have coupling relationships with other joints, making them of secondary importance, for example, the DIP of the fingers \cite{pip}. 

%

\begin{figure*}[!t]
    \includegraphics[width=\textwidth]{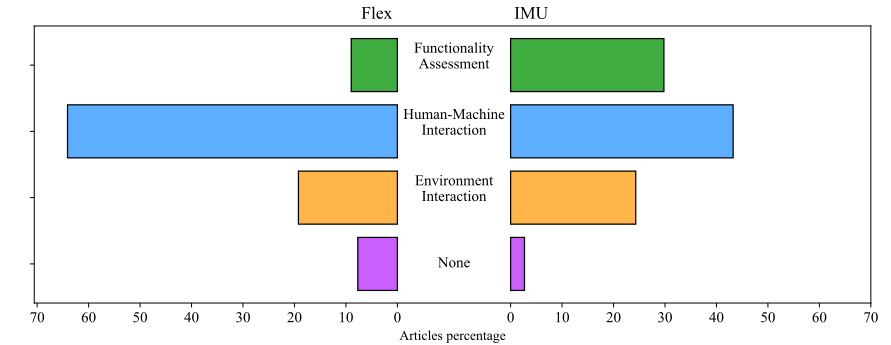}  \centering
  \caption{Comparison between the number of articles for each application field over the two main categories of tracking sensors: IMUs and flex sensors. Due to the imbalance of these two categories, the presented data have been normalized on the number of articles.}
  \label{fig:compa}
\end{figure*}

\begin{figure}[!t]
  \includegraphics[width = 0.7\columnwidth]{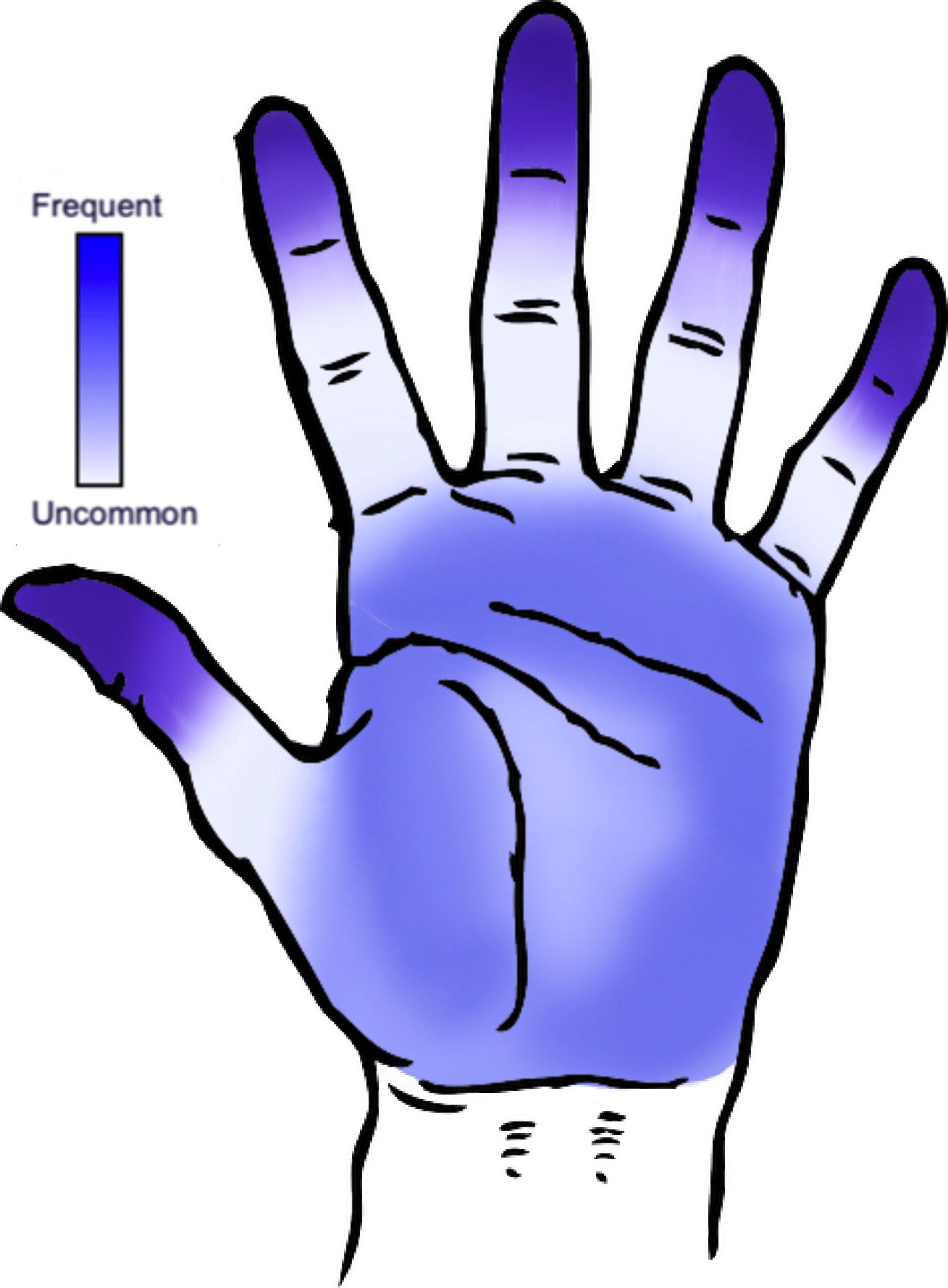}
  \centering
  \caption{The layout of pressure sensors on the surface of the hand. Darker colours imply that that area is chosen more frequently to place the sensors.}
  \label{fig:asadas}
\end{figure}

Fig. \ref{fig:dofdist} visually represents the literature focus dedicated to each hand joint, i.e., the area of the circle is proportional to the number of articles considering the related joint. 
Studies on fingers' PIP and thumb's DIP are by far the most common, that is, studies on MCP are around half those on PIP joints.
Instead, although it is essential for manipulation purposes, the most overlooked joint in the literature is the thumb TM. 
This observation can be explained by the difficulties in placing sensors on this point of the kinematic chain. 
On the contrary, placing sensors on fingers PIP and thumb DIP joints is easier, i.e., they have only one DoF, the related links are big enough to accommodate sensors, and they are characterized by a great range of motion unlocking simple studies on hand opening and closure.

In Fig. \ref{fig:applications}, we present the distribution of articles over the four application fields introduced in \ref{sec:Relevance of the Topic}. 
Human-Machine Interaction (54.7\%) is the most popular application field for custom data gloves, followed by Environment Interaction (20.7\%) and Functionality Assessment (17.9\%). 
Finally, there is a 6.7\% of articles without a specific application field. 
On the one hand, considering the distribution of sensor type over the four application fields (see Fig. \ref{fig:compa}), we notice that IMUs are more popular than flex sensors for Functionality Assessment. 
On the other hand, Human-Machine Interaction studies prefer flex sensors. 
This result could be linked with the previous observations about the number of DoFs that various sensing technologies can track.
Functionality Assessment includes several clinical-related applications such as detection and prevention of neurodegenerative diseases \cite{1213645, abtahi2020merging, anbalagan2022novel, sadhu2022towards}, analysis of patient improvements in physical therapy treatments \cite{kim2020wearable, sabry2020toward, Schuster-Amft, tavares2016data} and hand kinematic assessment \cite{libor, 13654}. Some applications focus on detecting recurrent patterns such as tremors and motor synergies. While IMUs are generally more widely used due to their ability to track higher degrees of freedom (DOFs), as illustrated in Fig. \ref{fig:applications}, careful sensor placement can make flex sensors equally effective. In contrast, applications such as hand kinematic assessment are closely tied to the kinematic model of the hand and inherently require extensive tracking capabilities. In this case, gloves based on inertial sensors appear to be the most appropriate choice, although it is still possible to consider a simplified kinematic model and use flex sensors.

Opposing considerations apply to Human-Machine Interaction, whose applications mainly include gesture-based communication \cite{liu2016interactive, informatics5020028, kanokoda2019gesture} and teleoperation \cite{chauhan2019grasp, choi2021development, yu2022end}.
In this case, the difference in the number of articles favours flex sensors as gesture recognition applications generally consider a limited set of hand joints (see Fig. \ref{fig:dofdist}) or directly the raw sensor data.

Regarding Environment Interaction, there are no significant differences in the number of articles for the two types of sensors. These applications focus on identifying and measuring the contact forces during the manipulation of objects \cite{komi, wanghyuk, boruah2022shape}. Therefore, the choice between the two tracking modes is likely to have less impact than the previous two categories.

Finally, None collects articles without a clear application of use, and it is greatly underrepresented compared to the other categories. Most of the articles we have classified in None relate to the presentation of new hardware architectures or the design of new types of sensors \cite{zhang2020ultrasensitive, vu2020highly, rand}. More in detail, inertial sensors consistently rely on MEMS technology, whereas research in flex sensors is more actively exploring the use of different bendable conductive materials. Consequently, we can observe a higher prevalence of flex gloves in the None category for these reasons.


Finally, we examined the distribution and total area covered by pressure sensors on the hand. 
The findings, shown in Fig. \ref{fig:asadas}, revealed that the most common location for pressure sensors is the fingertips, followed by the palm and the middle phalanges of the fingers. 
This result makes sense as the fingertips are essential for object manipulation, whereas the palm helps improve grip and generate friction when applying forces. 
The middle and proximal phalanges of the fingers are the least common choice and are mainly relevant for specific grasping tasks where the hand can wrap around a tool or for handling large, heavy objects. 
The size of the area covered by pressure sensors is also fundamental in glove design. 
We calculated the total sensorized area of each glove, taking into account the number and size of the sensors. 
The results showed a minimum area of approximately {\SI{1.8}{\centi\metre\squared}}\cite{marechal2009first} (excluding gloves with only one sensor \cite{steffen2020wearable, kerner2022wearable}) and a maximum of {\SI{172}{\centi\metre\squared}}. 
The latter value is an outlier as the corresponding glove is the only one using a grid-based sensor produced by Tekscan Inc.
Specifically, this glove employed five rectangular grids for the fingers and a larger square grid for the palm \cite{tekscan}. 
All other gloves used single FSR sensors with smaller active areas, resulting in a maximum value of {\SI{16.10}{\centi\metre\squared}} and an average of {\SI{8.46}{\centi\metre\squared}}. 
Considering that the average surface of a human palm is {\SI[multi-part-units = single]{217\pm19}{\centi\metre\squared}} for men and {\SI[multi-part-units = single]{189\pm21}{\centi\metre\squared}} for women \cite{lee2007determination}, the hand palm area usually considered for sensorization is a small portion of the overall palm.

The table provided in the supplementary material of this article presents the results of our literature review in chronological order.
Each row includes the article reference, year of publication, type of sensors embedded in the glove, range of tracked DoFs, palm area sensorized in millimetres, and the application category. 

\section{Conclusion}  
\label{sec:Conclusion}

Sensorized data gloves are a valuable tool for studying the interactions between hands and the environment. They have the potential to provide valuable insights into motion analysis and tactile perception; the proof of this is the rapid growth in the number of articles on the topic, which reflects the growing academic and industrial interest driven by the human-machine interaction application fields (see Fig. \ref{fig:narticles} and \ref{fig:applications}).
Furthermore, they are an effective alternative to vision-based methods for motion tracking, and the only available method for detecting tactile phenomena. 
While commercial solutions exist, our analysis is focused on custom gloves, which offer a high level of flexibility.

Sensors represent the most critical aspect of custom data gloves. 
In this review, we distinguish between sensors for motion detection and those for the detection of touch and contacts, and we analyze their use in the literature. 
Additionally, our analysis presented the three types of sensors used for motion tracking (i.e. flex sensors, IMUs and magnetic trackers) and explained their functioning.
However, we decided to focus most of our analysis on flex sensors and IMUs since magnetic trackers are underrepresented in the literature. Our comparison showed that flex sensors are more popular than inertial sensors, with approximately twice as many articles. This difference seems justified by the greater ease of use guaranteed by the first type of sensor that, once placed on the joint, only requires a quick calibration process. On the other hand, flex sensors can only track a single degree of freedom, which makes them unsuitable for tracking more complex joints. Moreover, it is difficult to place on a glove a sufficient number of flex sensors to trace the entire kinematic structure of the hand due to their size and inter-joint crosstalk. Therefore, inertial sensors are the most suitable choice for greater accuracy in the kinematic model and higher number of degrees of freedom.

Obviously, sensing limitations can influence the choice of joints to be tracked. 
Hand joints most commonly studied with data gloves are the fingers PIP and the thumb DIP due to their ease of sensorization. 
On the contrary, the TM joint of the thumb, which is fundamental for the opposition movement, is often neglected due to the difficulties in its tracking. 
Finally, tactile sensing is typically achieved in data gloves by mounting pressure sensors, for example, FSRs, on the palm and fingertips. 
FSRs can detect pressure intensity, but not the direction of exerted force. A complete sensorization of the hand requires sensor arrays, which can significantly increase the overall bulkiness of the glove.

More recently, new technologies for sensing tactile phenomena on large surfaces, such as robot skin, have been introduced in the literature. These techniques mimic the functioning of human skin by sensing and localizing contact forces using distributed sensors. However, they have not yet been implemented in data glove manufacturing. The cause of this gap is attributable to the difficult integration of this type of sensor, which would require the development of taxels of extremely small size or flexible enough to adapt to the heterogeneous surface of the fingers. Additionally, the palm is the most sensitive part of the hand, and covering its surface with robotic skin would almost completely obstruct the sense of touch and reduce the range of motion of the fingers, thus hindering the manipulation of objects.

We argue that, in the near future, the main challenges in developing custom data gloves will be related to improving the reproducibility of their design and increasing the level of customization of their components. 
To make these devices last longer and easier to replicate, it is essential for new research to share detailed information about the employed hardware.
Additionally, adopting a modular design approach could allow researchers to customize their gloves to fit the needs of specific applications rather than starting from scratch for each new project. 
Finally, we noticed a widespread lack of experimental validation of custom data gloves. 
This problem should be tackled by new research proposing objective protocols.

\section{Acknowledgment}
This research was made, in part, with the Italian government support under the National Recovery and Resilience Plan (NRRP), Mission 4, Component 2 Investment 1.5, funded from the European Union NextGenerationEU and awarded by the Italian Ministry of University and Research. This work was also supported by the CHIST-ERA (2014-2020) project InDex and received funding from the Italian Ministry of Education and Research (MIUR).



\ifCLASSOPTIONcaptionsoff
  \newpage
\fi

\bibliographystyle{IEEEtran} 
\bibliography{references.bib}





\end{document}